\documentclass[aps,prb,reprint,superscriptaddress]{revtex4-1}
\usepackage[english]{babel}
\usepackage[T1]{fontenc}
\usepackage[utf8]{inputenc}
\usepackage{natbib}

\usepackage{bm}
\usepackage{xcolor}
\usepackage{tabularx}
\usepackage{amssymb}
\usepackage{amsmath}
\usepackage{graphicx}
\usepackage{multirow}
\usepackage[pdftex]{hyperref}
\usepackage{lipsum}
\usepackage{soul}

\newcommand{\imag}{\mathrm{i}}

\begin{document}

\title{Fluctuations and noise-limited sensing near the exceptional\\ point of $\mathcal{PT}$-symmetric resonator systems}

\author{N. Asger Mortensen}
\email{asger@mailaps.org}
\affiliation{Center for Nano Optics, University of Southern Denmark, Campusvej 55, DK-5230 Odense M, Denmark}
\affiliation{Danish Institute for Advanced Study, University of Southern Denmark, Campusvej 55, DK-5230 Odense M, Denmark}
\affiliation{Center for Nanostructured Graphene, Technical University of Denmark, DK-2800 Kongens~Lyngby, Denmark}

\author{P.~A.~D. Gon\c{c}alves}
\affiliation{Center for Nano Optics, University of Southern Denmark, Campusvej 55, DK-5230 Odense M, Denmark}
\affiliation{Center for Nanostructured Graphene, Technical University of Denmark, DK-2800 Kongens~Lyngby, Denmark}
\affiliation{Department of Photonics Engineering, Technical University of Denmark, DK-2800 Kongens~Lyngby, Denmark}

\author{Mercedeh Khajavikhan}
\affiliation{CREOL/College of Optics and Photonics, University of Central Florida, Orlando, Florida 32816, USA}

\author{Demetrios N. Christodoulides}
\affiliation{CREOL/College of Optics and Photonics, University of Central Florida, Orlando, Florida 32816, USA}

\author{C.~Tserkezis}
\affiliation{Center for Nano Optics, University of Southern Denmark, Campusvej 55, DK-5230 Odense M, Denmark}

\author{C. Wolff}
\affiliation{Center for Nano Optics, University of Southern Denmark, Campusvej 55, DK-5230 Odense M, Denmark}

\date{\today}

\begin{abstract}
We theoretically explore the role of mesoscopic fluctuations and noise on the spectral and temporal properties of systems of $\mathcal{PT}$-symmetric coupled gain-loss resonators operating near the exceptional point, where eigenvalues and eigenvectors coalesce. We show that the inevitable detuning in the frequencies of the uncoupled resonators leads to an unavoidable modification of the conditions for reaching the exceptional point, while, as this point is approached in ensembles of resonator pairs, statistical averaging significantly smears the spectral features. We also discuss how these fluctuations affect the sensitivity of sensors based on coupled $\mathcal{PT}$-symmetric resonators. Finally, we show that temporal fluctuations in the detuning and gain of these sensors lead to a quadratic growth of the optical power in time, thus implying that maintaining operation at the exceptional point over a long period can be rather challenging. Our theoretical analysis clarifies issues central to the realization of $\mathcal{PT}$-symmetric devices, and should facilitate future experimental work in the field.
\end{abstract}

\pacs{Valid PACS appear here}
      
\maketitle

Pronounced sample-to-sample fluctuations constitute a hallmark of mesoscopic physics~\cite{Altshuler:1991}, where the finite number of degrees of freedom limits the self-averaging common to macroscopic systems. In mesoscopic systems the interactions of waves with disordered potentials lead to many fascinating phenomena~\cite{Beenakker}, including Anderson localization~\cite{Anderson:1958}, weak localization~\cite{Altshuler:1980}, and universal conductance fluctuations~\cite{Lee:1985}. A typical playground for such effects is many-body electron physics, which is rich on mesoscopic fluctuations~\cite{Narozhny:2000,Mortensen:2001,Lunde:2006,Goorden:2007}. Another class of interesting systems can be found in optics, with intriguing examples including random lasers~\cite{Lawandy:1994}, quantum optical entanglement in multiple-scattering media~\cite{Ott:2010}, as well as cavity-quantum electrodynamics~\cite{Sapienza:2010} and nanolasing~\cite{Liu:2014} with Anderson localized states. Traditionally, many mesoscopic wave-interference phenomena have been explored using the tight-binding model of condensed-matter physics~\cite{Datta:1995}, while its optical analog -- coupled-mode theory (CMT)~\cite{Fan:2003} -- has fostered the exploration of systems consisting of coupled resonators, with an emphasis on long chains (waveguides)~\cite{Yariv:1999}, and the rich interplay of slow-light phenomena with the presence of both loss and gain~\cite{Grgic:2012} as well as disorder-induced Anderson localization~\cite{Grgic:2011}. While the quantum dynamics is commonly governed by Hermitian equations of motion, the electrodynamics of optical systems is in general non-Hermitian due to the inevitable presence of material absorption, but also the possibility of introducing optical gain. However, recent years have witnessed not only efforts to realize loss-compensation in optical metamaterials~\cite{Xiao:2010}, but also the possibility to enable $\mathcal{PT}$-symmetric systems~\cite{El-Ganainy:2018,Longhi:2017}, where eigenvalues can be real despite the non-Hermitian aspects of the governing equations~\cite{Bender:1998}.

\begin{figure}[b!]
\centerline{\includegraphics*[width=1\columnwidth]{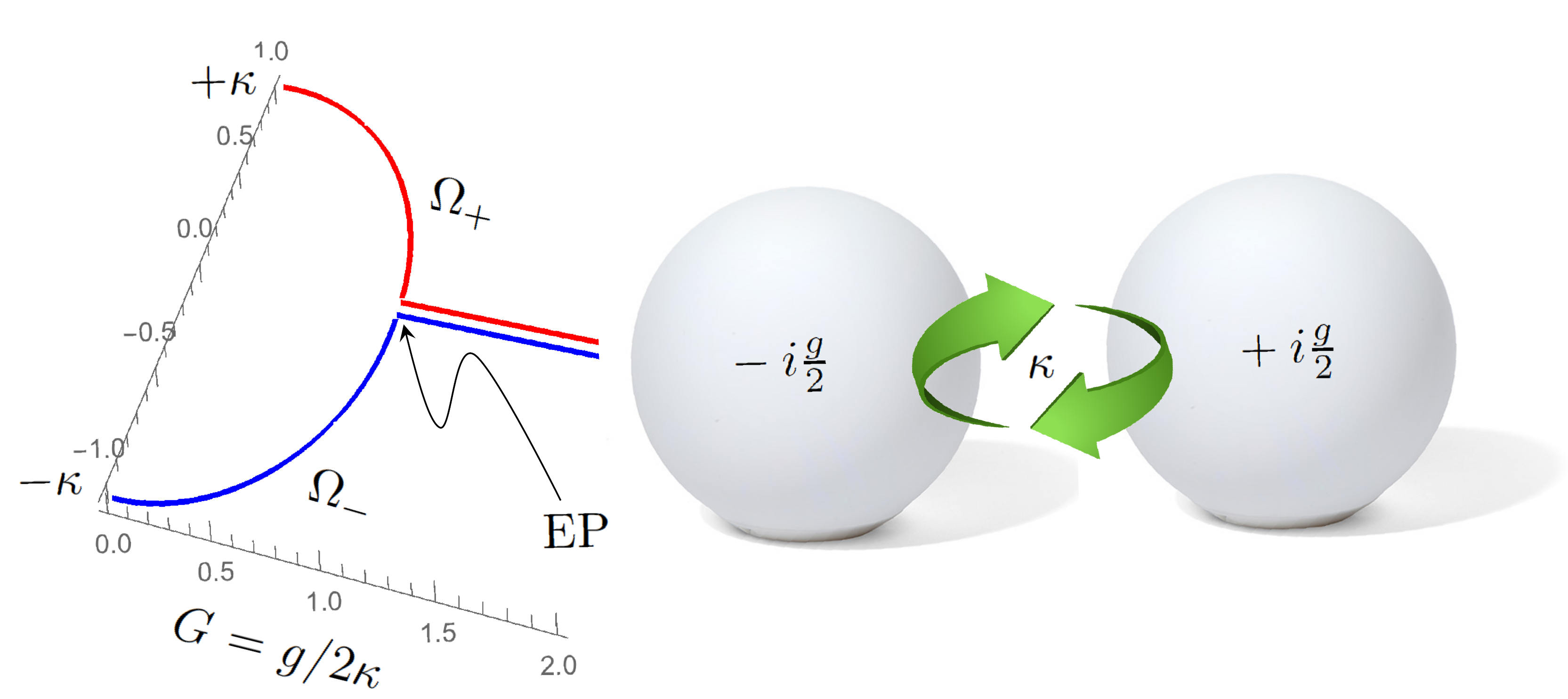}}
\caption{Schematic illustration of a $\mathcal{PT}$-symmetric dimer formed by two identical (no frequency detuning) coupled optical resonators, but with opposite values of the gain/loss parameter $G (=2g/\kappa)$. In the absence of gain/loss ($G=0$), the two resonators form common hybridized states with splitting by $2\kappa$, while for the exceptional point (EP) at $G=1$, the system is degenerate, i.e. $\Omega_+=\Omega_-$, with coalescing eigenstates.
}\label{fig1}
\end{figure}

Here, we turn to finite $\mathcal{PT}$-symmetric systems and illustrate interesting new mesoscopic fluctuations of the spectral properties near the exceptional point (EP) with coalescing eigenstates. For the transparency of our illustration, we consider a problem of two coupled resonators as illustrated in Fig.~\ref{fig1}. Within CMT, the dynamics is governed by a Schr\"odinger-like equation
\begin{align}\label{eq:EQM}
  \imag \partial_t 
  \underbrace{ \begin{pmatrix} a(t) \\ b(t) \end{pmatrix} }_{ \psi(t) }
  = & 
  \underbrace{ \begin{pmatrix}
    \omega_a-\imag \frac{g}{2}& \kappa \\ \kappa & \omega_b+\imag \frac{g}{2}
  \end{pmatrix} }_{ \mathcal{H} }
  \begin{pmatrix}
    a(t) \\ b(t)
  \end{pmatrix},
\end{align}
where $\omega_a$, $\omega_b$, $a$ and $b$ are the resonance frequencies of the 
uncoupled resonators and the amplitudes of their respective modes, $\kappa$ is 
the coupling parameter (which can be chosen real-valued), and $g$ 
characterizes the gain and damping of the two resonators.
For convenience we have introduced symbols $\psi(t)$ and $\mathcal{H}$ for 
the state and the Hamiltonian, respectively.
In the analysis of such a problem, it is customary to study the stationary
solutions (we will come back to the time-evolution towards the end of this paper), i.e. the eigenvalue problem $\mathcal{H}\psi=\omega\psi$. For $g=0$, this constitutes a Hermitian problem and corresponds 
to the usual hybridization of two levels, i.e., with the bonding and 
anti-bonding states (notation inherited from molecular orbital bonding theory) 
having real eigenfrequencies and in the $\omega_a=\omega_b$ case being 
separated by an energy of $2\kappa$. 

In the presence of a finite $g$, the system is non-Hermitian, while $\mathcal{PT}$-symmetry may still allow real-valued eigenfrequencies~\cite{Bender:1998}, depending on the strength of the gain $g$ relative to the coupling $\kappa$. Perhaps the most notable characteristic of a $\mathcal{PT}$-symmetric system is a $\mathcal{PT}$-symmetry breaking transition that takes place around $g/2\kappa=1$. In optical settings, this abrupt phase transition has been experimentally demonstrated in coupled waveguides and cavities, by measuring both the real and imaginary components of the eigenvalues, as well as by observing the evolution of the corresponding mode profile~\cite{Guo:2009,Rueter:2010,Regensburger:2012,Brandstetter:2014,Peng:2014,Hodaei:2014}.

In order to analyze the influence of temporally fluctuating environments or sample-to-sample fluctuations associated with inevitable small variations in $\omega_a$ and $\omega_b$, we shall in the following allow a small, but finite frequency detuning between the two coupled resonators.  
To ease our notation, we first define a normalized frequency $\Omega=\omega/\kappa$ and center frequency $\bar{\Omega}=(\omega_a+\omega_b)/2\kappa$, while the normalized detuning of the two resonances is denoted by $\Delta=(\omega_a-\omega_b)/2\kappa$. The eigenvalue problem $\mathcal{H}\psi=\omega\psi$ now takes the form
\begin{subequations}
\begin{equation}\label{eq:eqm-normalized}
  \begin{pmatrix}
    \bar\Omega+\Delta-\imag G& 1 \\ 1 & \bar\Omega-\Delta+\imag G
  \end{pmatrix}\begin{pmatrix}
    a \\ b
  \end{pmatrix}=\Omega \begin{pmatrix}
    a \\ b
  \end{pmatrix}
\end{equation}
where $G=g/2\kappa$ is the normalized parameter central to the analysis of exceptional points in this problem. By straightforward diagonalizing we get
\begin{equation}\label{eq:spectrum}
\Omega_{\pm}=\bar{\Omega}\pm \sqrt{1-(G+\imag  \Delta)^2 },
\end{equation}
with corresponding eigenvectors
\begin{equation}
\psi_\pm= \begin{pmatrix}
a \\ b
\end{pmatrix}_{\pm}= \begin{pmatrix}
-\imag (G+\imag\Delta) \pm\sqrt{1-(G+\imag \Delta )^2} \\ 1
\end{pmatrix}.
  \label{eqn:eigenstates}
\end{equation}
\end{subequations}
Obviously, the eigenfrequencies of the coupled system can in general be complex, i.e. $\Omega=\Omega'+\imag\Omega''$ and we immediately see how detuning enters simply as an imaginary part of the gain parameter: $G^2\rightarrow (G+\imag \Delta )^2$. Eqs.~(\ref{eq:spectrum},\ref{eqn:eigenstates}) nicely illustrate how both the eigenvalues and the eigenstates coalesce ($\Omega_+=\Omega_-$ and $\psi_+=\psi_-$) when the square root vanishes, forming an exceptional point.
If the two un-coupled resonators are perfectly aligned ($\Delta=0$), this occurs for $G=1$, where the gain and loss is exactly balanced by an appropriate coupling constant. 

Under realistic experimental conditions, the built-in material loss can always be compensated by carefully adjusting the gain, e.g., through electrical pumping of one of the resonators~\cite{Gao:2017}. However, no matter the efforts spent in fabricating resonators with similar resonance frequencies, there is always some small, yet inevitable frequency detuning. Moreover, this detuning will vary from sample to sample. In a particular sample the detuning is also likely to fluctuate over time due to unavoidable fluctuations of the environment. In this paper, we aim to study the interplay of such sample-to-sample fluctuations and its behavior and magnitude near exceptional points. We also consider possible implications of fluctuating environments for the exploration of exceptional points in sensing~\cite{Miller:2017}.

\emph{Below the exceptional point.} For low gain ($G\ll 1$), below the exceptional point, we have to leading order in the detuning that
\begin{subequations}
\begin{eqnarray}
\Omega'_{\pm}&\simeq& \bar{\Omega}\pm (1+\tfrac{1}{2}\Delta^2),\\
\Omega''_{\pm}&\simeq& \mp G \Delta.
\end{eqnarray}
\end{subequations}

In the ideal case ($\Delta=0$), this regime is characterized by a real-valued spectrum, i.e. $\Omega=\bar{\Omega}\pm 1$. However, for a small, but finite detuning, the imaginary part is finite despite the symmetric gain/loss arrangement. In other words, the finite detuning breaks the $\mathcal{PT}$-symmetry associated with perfectly aligned resonators. This is also immediately clear by noticing that the Hamiltonian does not equal its adjoint, i.e. $\mathcal{H}\neq \mathcal{H}^\dagger$.


\begin{figure}[t!]
\centerline{\includegraphics*[width=1\columnwidth]{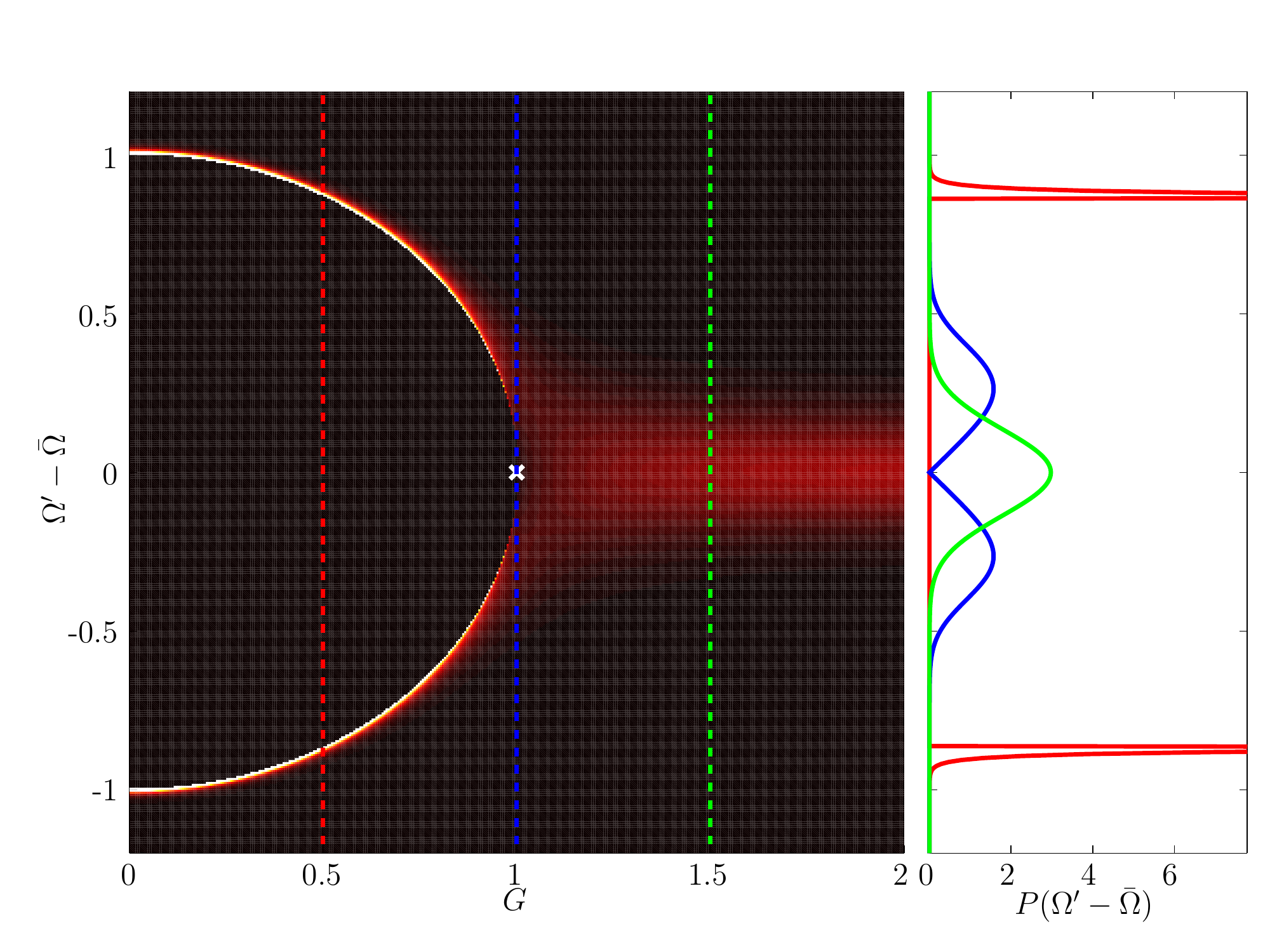}}
\centerline{\includegraphics*[width=1\columnwidth]{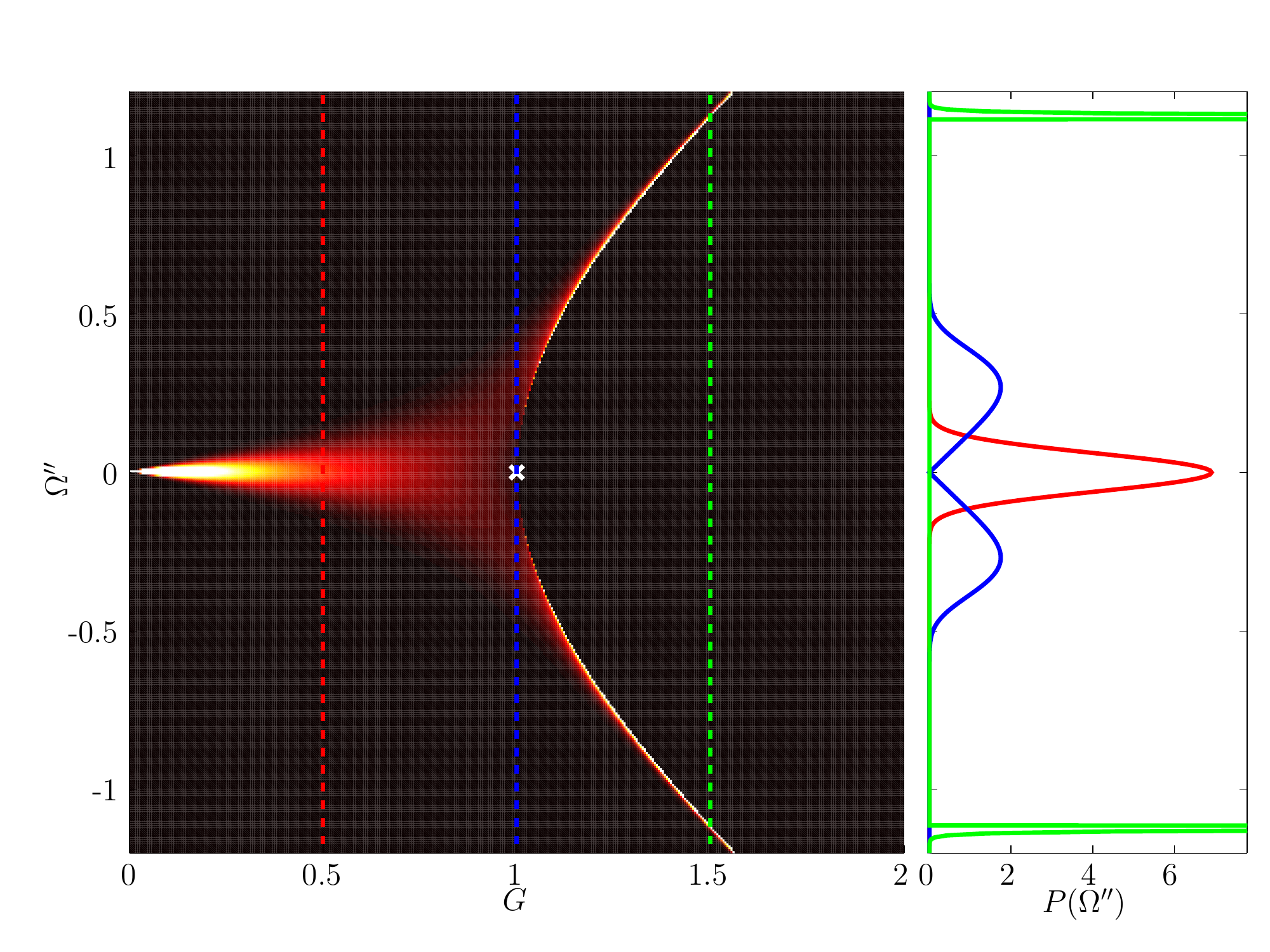}}
\caption{Plots of the distribution of complex eigenfrequencies $\Omega=\Omega'+\imag\Omega''$ for varying $G$. The upper panel shows $P(\Omega')$ while $P(\Omega'')$ is displayed in the lower panel for an ensemble of coupled resonators with $\sigma=0.1$. Clearly, the sample-to-sample fluctuations are pronounced as one approaches the exceptional point.
}\label{fig2}
\end{figure}

\emph{Fluctuations near the exceptional point.} 
In order to see that detuning changes the conditions for having an exceptional point, we expand the exact expression [Eq.~\eqref{eq:spectrum}] around the exceptional point $G=1$; to leading order in $\Delta$ we 
get\cite{Pick:2017}
\begin{equation}
  \Omega_{\pm}\simeq \bar{\Omega}\pm 
  (1-\imag)\sqrt{\Delta}.
  \label{eq:leading_order_correction}
\end{equation}
The detuning lifts the degeneracy that would otherwise be associated with the exceptional point of two perfectly aligned resonators ($\Delta=0$).

Away from the exceptional point, systems are commonly affected linearly by perturbation. However, the fact that the splitting scales as $\sqrt{\Delta}$ is an interesting manifestation of the system being very susceptible to perturbations near the exceptional points~\cite{Wiersig:2014,Hodaei:2017}. Obviously, this can be used to our advantage in the context of optical sensors~\cite{Miller:2017}, but has the natural drawback that the system is also very sensitive to any undesired, yet practically inevitable degrees of freedom associated with fabrication imperfections or fluctuating environments (e.g. temperature shifts or noise in the gain parameter).

We now assume an ensemble of resonator pairs with a Gaussian distribution of the
detuning parameter
\begin{align}\label{eq:P0}
  P_0(\Delta) = \frac{1}{\sqrt{2\pi} \sigma} \exp(-\tfrac{1}{2}\Delta^2/\sigma^{2}).
\end{align}
This can be interpreted either as fabrication tolerance or as temporal 
fluctuations assuming that an ergodic approximation to the system dynamics is
valid.

In order to appreciate the dramatic effect this has on the spectrum especially
near the exceptional point, we show in Fig.~\ref{fig2} the distribution of the 
eigenvalues' real and imaginary parts where the variance $\sigma=0.1$ was 
chosen sufficiently small that the common regime with $G=0$ is only slightly 
broadened. 
However, in the vicinity of the exceptional point (and beyond), we observe a 
very pronounced smearing of the spectral features. 

\begin{figure}[b!]
\centerline{\includegraphics*[width=1\columnwidth]{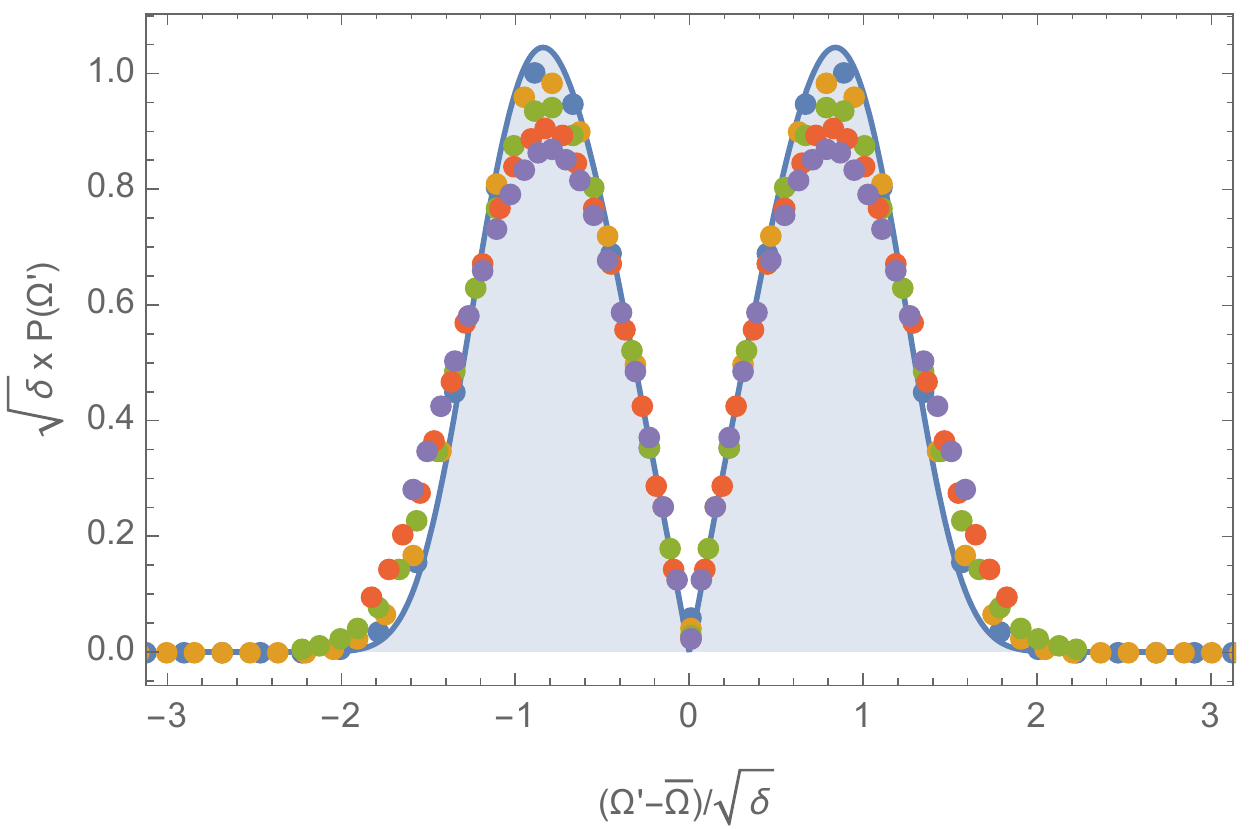}}
\caption{Plot of the distribution of $P(\Omega')$ versus $\Omega'-\bar\Omega$ at the exceptional point ($G=1$) for ensembles of coupled resonators with $\sigma=0.05, 0.1, 0.2, 0.3$, and $0.4$. Data points are the results of numerical ensemble averaging of the spectra associated with Eq.~\eqref{eq:spectrum}, while the filled curve shows the approximate universal result from Eq.~\eqref{P_approx}.
}\label{fig3}
\end{figure}

Pursuing a deeper understanding of this numerical observation, we 
proceed with analytical calculations based on the leading-order correction 
Eq.~\eqref{eq:leading_order_correction}.
The eigenvalue $\Omega$ is not a convenient quantity to study at an isolated 
point of degeneracy that is lifted by a statistical process. 
Instead, we shall focus on the splitting of the eigenvalue's real part
\begin{align}
  \Sigma = \Omega'_+ - \Omega'_- \approx 2 \sqrt{|\Delta|}.
\end{align}
It should be noted that corresponding expressions for fluctuations in the gain 
coefficient and for the splitting of $\Omega''$ are very similar.
Its ensemble average is
\begin{align}
  \left<\Sigma\right> = & 2\int_{-\infty}^{\infty} {\rm d}\Delta\, \sqrt{|\Delta|}P(\Delta)\nonumber
  \\
  = & 2^{5/4} \Gamma \left(\frac{3}{4}\right)\sqrt{\frac{\sigma}{\pi}}
  \approx 1.64 \sqrt{\sigma}.
\end{align}
Given the Gaussian distribution of the detuning, the distribution of detunings at the exceptional point can now be evaluated (see appendix~\ref{appandix:A})
\begin{eqnarray}
  P(\Omega') \simeq 
  \frac{1}{\sqrt{\sigma}}
  F\left(\frac{\Omega' -\bar{\Omega}}{\sqrt{ \sigma}}\right)\label{P_approx}
\end{eqnarray}
where 
$F(x)=\left(\tfrac{8}{\pi}\right)^{1/2} |x| \exp\left(-\tfrac{1}{2}x^4\right)$. 
This approximate universal distribution shown in Fig.~\ref{fig3} illustrates an interesting ensemble-averaged broadening of levels 
inside the gap, i.e. a $P(\Omega') \propto |\Omega' -\bar{\Omega}|$ for 
energies smaller than the detuning. 
Within the square root scaling law, $P(\Omega'')$ is distributed in the same 
manner (see appendix~\ref{appandix:A}); the cuts through the exceptional point (solid blue
curves) of the two panels in Fig.~\ref{fig2} are nearly identical, but not 
quite due to the finite $\sigma$.

\begin{figure}[t!]
\centerline{\includegraphics*[width=1\columnwidth]{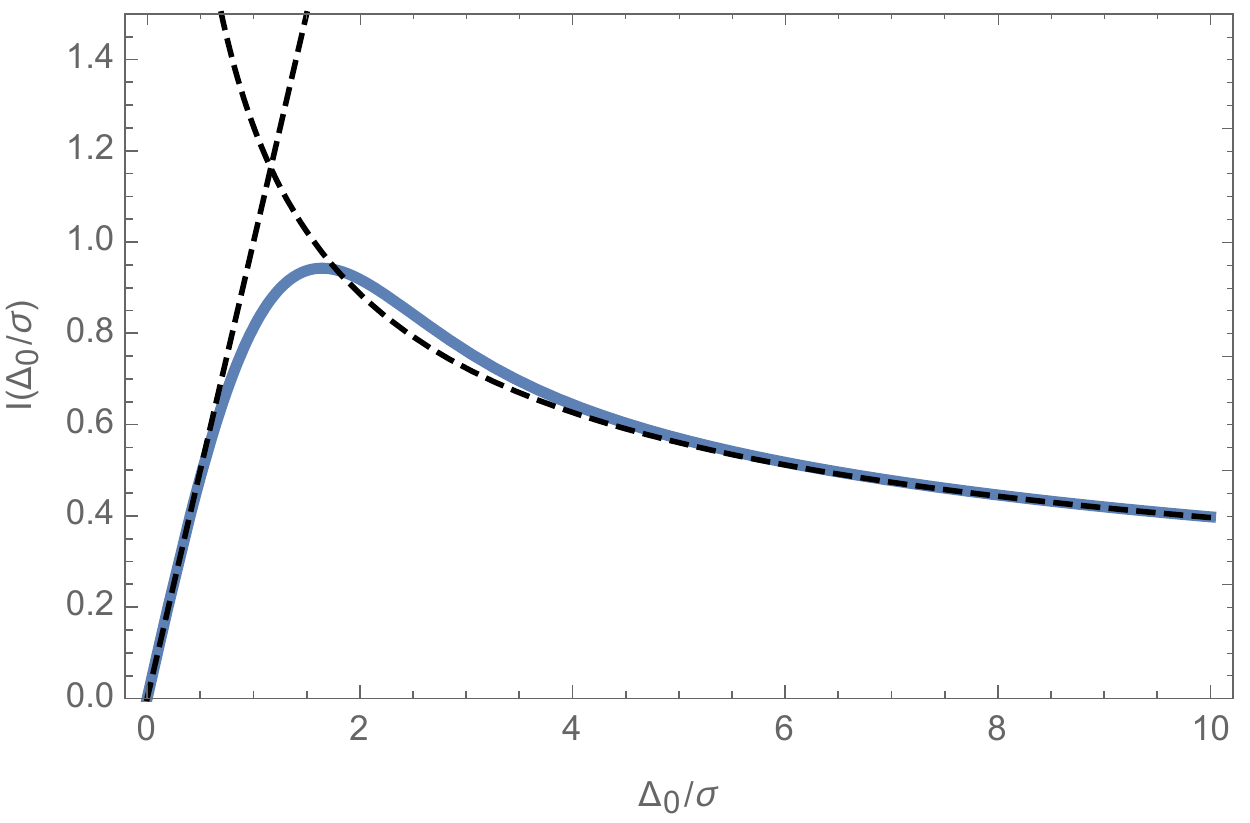}}
\caption{Plot of the integral $I(\Delta_0/\sigma)$ in Eq.~\eqref{eq:sensitivity} versus $\Delta_0/\sigma$ (solid line) with the asymptotic behaviors in Eq.~\eqref{I_approx} indicated by dashed lines.
}\label{fig4}
\end{figure}

\emph{Sensitivity of fluctuating sensors.} 
It is not entirely surprising that statistical detuning leads to a non-zero
average eigenvalue splitting.
The natural next question is how this affects the performance of a sensor, i.e.
how the average splitting $\langle \Sigma \rangle$ reflects an additional,
non-fluctuating detuning.
We now assume that this detuning parameter has two contributions: firstly a 
fluctuating detuning due to unintended noise, which is inevitably present in 
any realization of such systems, and secondly the signal $\Delta_0$ that is 
meant to be detected or sensed. For the detuning probability distribution
\begin{align}\label{eq:P-shifted}
  P(\Delta) = \frac{1}{\sqrt{2 \pi} \sigma} \exp[-\tfrac{1}{2}(\Delta - \Delta_0)^2/ \sigma^{2}],
\end{align}
the sensitivity of the time-averaged frequency splitting can now be written as (see appendix~\ref{appandix:B})
\begin{equation}
\frac{\partial \big< \Sigma\big>}{\partial\Delta_0}
=\sqrt{\tfrac{2}{\pi}}\sigma^{-1/2}\underbrace{\int_{-\infty}^\infty {\rm d}x\, \sqrt{\left|x+\tfrac{\Delta_0}{\sigma}\right|}x\exp\left(-\tfrac{1}{2}x^2\right)}_{I(\Delta_0/\sigma)}.\label{eq:sensitivity}
\end{equation}
Here, the integral can be approximated in the small and large-signal limits

\begin{equation}\label{I_approx}
I(\Delta_0/\sigma) \approx \left\{\begin{matrix}
\frac{\Delta_0}{\sigma} &,& \Delta_0\ll \sigma \\\\
\sqrt{\frac{\pi}{2}}\sqrt{\frac{\sigma}{\Delta_0}} &,& \Delta_0\gg \sigma\end{matrix}\right.
\end{equation}
and in Fig.~\ref{fig4} we show these asymptotic behaviors along with a full numerical evaluation of the integral. The integral is always smaller than unity, implying that the sensitivity is noise limited, i.e. $\partial\big< \Sigma\big>/\partial\Delta_0< \sigma^{-1/2}$. The sensitivity should be contrasted to the case in the absence of fluctuations, where there is a tremendous sensitivity to small signals, i.e. $\partial  \Sigma/\partial\Delta_0=\Delta_0^{-1/2}$. 
Indeed, from Eq.~\eqref{I_approx} we recover this result for $\sigma \rightarrow 0$. On the other hand, it is quite clear how the sensitivity vanishes linearly in the low-signal limit, where the perturbation is dressed by the noise.

\emph{Time evolution.} 
The use of exceptional points in highly sensitive sensors is seriously hampered
by the low-frequency tail of the temporal fluctuations of detuning and gain.
This tail is called drift and must be compensated by a feedback loop, i.e. the
sensor is kept at the exceptional point by constantly adjusting detuning and
pump power and the actually measured quantities are the values of these 
feedback variables (e.g. pump power for the gain or heating currents for the
detuning).
For this, it seems necessary to keep the sensor at the exceptional point over 
an extended period of time.
Naively, this seems trivial, because the eigenstate has a real eigenvalue and
one would therefore expect the time-evolution to be stationary and neither 
growing nor decaying in time.
In reality, this is not the case.
We now return to the equation of motion [Eq.~\eqref{eq:EQM}], which at the exceptional point reads
\begin{align}
  \imag \partial_\tau \psi = &
  \mathcal{H}_0 \psi; &
  \mathcal{H}_0 = &
  \begin{pmatrix} 
    \bar\Omega - \imag & 1 \\
    1 & \bar\Omega + \imag  
  \end{pmatrix},
\end{align}
where $\tau = \kappa t$ is the dimensionless time variable.
We can solve this formally using the time-evolution operator~\cite{Longhi:2018} 
\begin{align}
  \psi(\tau) 
  = & \mathcal{U}_0(\tau) \psi(0)
  = \exp(-\imag \mathcal{H}_0 \tau) \psi(0).
\end{align}
The matrix exponential can be simplified by decomposing
$\mathcal{H}_0 = \bar\Omega \mathbb{I} + A$, where $\mathbb{I}$ is the unit 
matrix and $A = \begin{pmatrix} -\imag & 1 \\ 1 & \imag \end{pmatrix}$.
Since $\mathbb{I}$ and $A$ commute, we find:
\begin{align}
  \mathcal{U}(\tau) 
  = & \exp(-\imag \bar\Omega \tau) 
  \sum_{n=0}^\infty \frac{(-\imag \tau)^n}{n!} A^n\nonumber
  \\
  = & \exp(-\imag\bar\Omega \tau) (\mathbb{I} - \imag A \tau),
  \label{eqn:time-evolution}
\end{align}
because $A^2$ vanishes ($A$ is nilpotent).
This dynamics is highly reminiscent of the critically damped classical
harmonic oscillator, whose time-evolution is a superposition of 
$h_1(t) \simeq \exp(-\gamma t)$ and $h_2(t) \simeq t\exp(-\gamma t)$.
Indeed, the critically damped harmonic oscillator formulated as two coupled
first-order differential equations results in an exceptional point in the
coupling matrix and constitutes a beautiful didactic example for this 
phenomenon~\cite{Dolfo:2018}.
Eq.~\eqref{eqn:time-evolution} has several important implications.
Firstly, it means that the overall optical power is not conserved when 
operating at an exceptional point. Instead, the optical amplitudes in general 
grow linearly and the optical energy therefore quadratically in time.
This makes it rather difficult to keep a sensor at an exceptional point for an
extended period of time.
Secondly, this demonstrates nicely that having a Hamiltonian with only real 
eigenvalues is not sufficient to ensure energy conservation~\cite{Bender:2007}. To address the former issue, in situations where high sensitivity is desired, one may encounter a daunting scenario of being in constant need for continuously monitoring and correcting the system so as to bring it back at the exceptional point. In practice, however, this task can be carried out in a considerably less demanding fashion by using the phase transition associated with the EP as a means to eliminate the requirement for constant correction. For example, by modulating the magnitude of the gain around the nominal value for exceptional point and by monitoring the output signal, one can determine the parameter range where the derivative of the response over time reaches its extremum~\cite{Ren:2017,Chen:2017}.


\emph{Discussion \& conclusion.} So far, we have discussed the classical electrodynamics at exceptional points of $\mathcal{PT}$-symmetric systems, where the spectrum can be real despite the presence of both loss and gain. We have emphasized mesoscopic fluctuations of a classical origin, while we speculate that also quantum optics and quantum fluctuations would experience a dramatic enhancement near the exceptional point. Quantum emitter dynamics in the presence of exceptional points is in itself interesting~\cite{Pick:2017}. In the present context, we note that loss-compensated metamaterials do not necessarily exhibit the dynamics of ideal loss-less structures when probed with quantum optics~\cite{Amooghorban:2013}, and as such there might also be interesting quantum fluctuation properties to be explored in the vicinity of exceptional points.

Focusing here on the role of mesoscopic fluctuations and noise on the spectral and temporal properties of systems of $\mathcal{PT}$-symmetric coupled gain-loss resonators operating near the exceptional point (EP), we have shown that the inevitable detuning in the frequencies of the uncoupled resonators leads to an unavoidable modification of the conditions for reaching the exceptional point. In ensembles of resonator pairs, statistical averaging significantly smears the spectral features which leaves sensitivity of EP-based sensors noise-limited. Finally, we have shown how temporal fluctuations in the detuning and gain of such sensors lead to a quadratic growth of the optical power in time, thus implying that maintaining operation at the exceptional point over a long period can be rather challenging. 
\\

\emph{Acknowledgments.} N.~A.~M. is a VILLUM Investigator supported by VILLUM Fonden (grant No. 16498). The Center for Nano Optics is financially supported by the University of Southern Denmark (SDU 2020 funding). The Center for Nanostructured Graphene is sponsored by the Danish National Research Foundation (Project No. DNRF103). C.~W. acknowledges funding from MULTIPLY  fellowships under the Marie Sk\l{}odowska-Curie COFUND Action (grant agreement No. 713694).

\begin{appendix}

\section{Distribution functions}
\label{appandix:A}
Knowing the distribution $P(\Delta)$ and the relation between $\Omega'$ and $\Delta$, the distribution $P(\Omega')$ can be calculated straightforwardly. In the following, we do this for a symmetric distribution $P_0(\Delta)$ with zero mean value,
\begin{eqnarray}
P(\Omega') &\simeq & \int_{-\infty}^\infty {\rm d}\Delta\, P_0(\Delta)\delta\left(\Omega'-{\rm Re}\{\bar{\Omega}\pm (1-\imag)\sqrt{\Delta}\}\right)\nonumber\\
&=&2\int_{0}^\infty {\rm d}\Delta\, P_0(\Delta)\delta\left(\Omega'-\bar{\Omega}\pm \sqrt{\Delta}\right).
\label{P_approx-app}
\end{eqnarray}
Performing the integral with Eq.~\eqref{eq:P0} we arrive at Eq.~\eqref{P_approx}. Note that $\int_{-\infty}^\infty {\rm d}\Omega'\,P(\Omega')=2$ due to inclusion of both $\Omega_\pm$ branches. In a similar way, it is perhaps no surprise that the imaginary part is distributed in the same way, i.e. for $P(\Omega'')$ we get

\begin{eqnarray}
P(\Omega'') &\simeq & \int_{-\infty}^\infty {\rm d}\Delta\, P_0(\Delta)\delta\left(\Omega''-{\rm Im}\{\pm (1-\imag)\sqrt{\Delta}\}\right)\nonumber\\
&=& 2\int_0^{\infty} {\rm d}\Delta\, P_0(\Delta)\delta\left(\Omega''\pm \sqrt{\Delta}\right).\label{P''_approx}
\end{eqnarray}
Comparing to Eq.~\eqref{P_approx-app} it is now immediately clear that $P(\Omega'')=\frac{1}{\sqrt{\sigma}}
F\left(\frac{\Omega''}{\sqrt{\sigma}}\right)$.

\section{Sensitivity to external signals}
\label{appandix:B}
We consider a sensing situation with a shifted Gaussian distribution $P(\Delta)$ centered around $\Delta_0$, see Eq.~\eqref{eq:P-shifted}. The mean value $\big< \Omega'\big>=\int d\Omega'\, \Omega' P(\Omega')$ now becomes

\begin{eqnarray}
\big< \Omega'_\pm\big>
&=&\int d\Omega'\, \Omega' \nonumber\\
&&\times\int_{-\infty}^\infty {\rm d}\Delta\, P(\Delta)\delta\left(\Omega'-{\rm Re}\{\bar{\Omega}\pm (1-\imag)\sqrt{\Delta}\}\right)\nonumber\\
&=&\int_{-\infty}^\infty {\rm d}\Delta\, {\rm Re}\{\bar{\Omega}\pm (1-\imag)\sqrt{\Delta}\}P(\Delta)
\end{eqnarray}
where we have performed the $\Omega'$ integral with the aid of the Dirac delta function. Next, we split the integral into its positive and negative parts in order to take the real part. Eventually, this gives

\begin{eqnarray}
\big< \Omega'_\pm\big>&=&\bar\Omega \pm \int_{-\infty}^\infty {\rm d}\Delta\, \sqrt{|\Delta|}P(\Delta)
\end{eqnarray}
Next, we turn to the splitting $\Sigma=\Omega'_+-\Omega'_-$ and calculate the sensitivity of this to the signal, i.e., 
\begin{eqnarray}
\frac{\partial \big< \Sigma\big>}{\partial\Delta_0}&=&\frac{2}{\sigma^2}\int_{-\infty}^\infty {\rm d}\Delta\, \sqrt{|\Delta|}(\Delta-\Delta_0)P(\Delta)
\end{eqnarray}
where we have used that $\frac{\partial}{\partial\Delta_0}P(\Delta)=\frac{\Delta-\Delta_0}{\sigma^2}P(\Delta)$, see Eq.~\eqref{eq:P-shifted}. To proceed, we make the substitution $x=(\Delta-\Delta_0)/\sigma$ which brings us to Eq.~\eqref{eq:sensitivity}.

\end{appendix}

\bibliography{ref}

\newpage

\end{document}